\documentclass[twocolumnappendix,numberedappendix,appendixfloats]{emulateapj}
\usepackage{natbib}
\usepackage{graphicx}
\usepackage{epsfig}
\usepackage{subfigure}
\usepackage{rotating}
\usepackage{lmodern}% http://ctan.org/pkg/lmodern
\usepackage{slantsc}% http://ctan.org/pkg/slantsc
\slugcomment{Submitted, 19-11-2015; Accepted 22-01-2016.}

\shorttitle{PAH emission in \textit{Spitzer} IRS maps}
\shortauthors{D. J. Stock et al.}

\begin{document}
\newcommand{\HII}{H~{\sc ii}}
\newcommand{\Gnaught}{G$_0$}
\newcommand{\cii}{[C~{\sc ii}]}
\newcommand{\oi}{[O~{\sc i}]}
\newcommand{\ci}{[C~{\sc i}]}

\title{Polycyclic Aromatic Hydrocarbon emission in \textit{Spitzer}/IRS maps I: \\ Catalog and simple diagnostics}

\author{D.~J.~Stock\altaffilmark{1,$\dagger$}, W.~D.-Y. Choi\altaffilmark{1}, L.~G.~V.~Moya\altaffilmark{1}, J. N. Otaguro\altaffilmark{1}, S.~ Sorkhou\altaffilmark{1},\\ L.~J.~Allamandola\altaffilmark{2},  A.~G.~G.~M.~Tielens\altaffilmark{3}, E.~Peeters\altaffilmark{1,4}}

\altaffiltext{1}{Department of Physics and Astronomy, University of Western Ontario, London, ON, N6A 3K7, Canada}
\altaffiltext{2}{NASA Ames Research Center, MS 245-6, Moffett Field, CA 94035-0001, USA}
\altaffiltext{3}{Leiden Observatory, Leiden University, PO Box 9513, 2300 RA, The Netherlands}
\altaffiltext{4}{SETI Institute, 189 Bernardo Avenue, Suite 100, Mountain View, CA 94043, USA}
\altaffiltext{$\dagger$}{Email: dstock4@uwo.ca}

\begin{abstract}
We present a sample of resolved galactic \HII\ regions and photodissociation regions (PDRs) observed with the \textit{Spitzer} infrared spectrograph (IRS) in spectral mapping mode between the wavelengths of 5--15 \micron. For each object we have spectral maps at a spatial resolution of $\sim$4\arcsec\ in which we have measured all of the mid-infrared emission and absorption features. These include the PAH emission bands, primarily at 6.2, 7.7, 8.6, 11.2 and 12.7 \micron, as well as the spectral emission lines of neon and sulfur and the absorption band caused by silicate dust at around 9.8 \micron. In this work we describe the data in detail, including the data reduction and measurement strategies, and subsequently present the PAH emission band intensity correlations for each of the objects and the sample as a whole. We find that there are distinct differences between the sources in the sample, with two main groups, the first comprising the \HII\ regions and the second the reflection nebulae (RNe). Three sources, the reflection nebula NGC~7023, the Horsehead nebula PDR (an interface between the \HII\ region IC~434 and the Orion B molecular cloud) and M 17, resist this categorization, with the Horsehead PDR points mimicking the RNe and the NGC~7023 fluxes displaying unique bifurcated appearance in our correlation plots. These discrepancies seem to be due to the very low radiation field experienced by the Horsehead PDR and the very clean separation between the PDR environment and a diffuse environment in the NGC~7023 observations.
\end{abstract}


\section{Introduction}

Over the past twenty years or so the sample of astronomical objects known to emit the mid-infrared (MIR) aromatic bands present from 3--20 \micron\ has increased dramatically, and new discoveries regarding the inter-relations between the different bands have been revealed (e.g., \citealt{1996ApJ...460L.119J, 2002A&A...390.1089P, 2004ApJ...611..928V, 2004ApJ...617L..65P, 2008ApJ...679..310G, 2012ApJ...747...44P, 2012A&A...542A..69P, 2013ApJ...769..117B, 2014ApJ...795..110B, 2014ApJ...791...99S, 2015ApJ...811..153S}, Peeters et al. 2016). 

The observed emission of the MIR bands is now commonly associated with the vibrational relaxation of ultraviolet-pumped polycyclic aromatic hydrocarbon (PAH) molecules, with several tools now available to interpret the MIR bands in terms of PAH emission (e.g., PAHFIT, \citealt{2007ApJ...656..770S}; the NASA Ames PAH database fitting routines, \citealt{2010ApJS..189..341B,2014ApJS..211....8B}; and PAHTAT, \citealt{2012A&A...542A..69P}). The PAH bands are observed in most astronomical contexts, particularly ones in which there is a significant UV radiation field (e.g., \HII\ regions, RNe and planetary nebulae, \citealt{1996A&A...315L.337V, 2001A&A...370.1030H}; galactic cirrus and dark cloud surfaces, \citealt{1996A&A...315L.353M}; external galaxies, \citealt{1998ApJ...498..579G}) leading to the conclusion that PAH molecules are ubiquitous in the Universe.% The life cycle of PAH molecules is still a matter of debate, with the leading theory being that they form in the slow, dense outflows of evolved, low mass, carbon rich stars \citep{1991ApJ...377..187L}. However, others such as \citet{2010A&A...510A..36M} have suggested that the lifetime of PAH molecules in the ISM is very short as the efficiency of their destruction in interstellar shocks is thought to be high. More recent lab-based studies have shown that the PAH molecules are actually more stable than previously thought, which will potentially resolve the key outstanding issue \citep{2015ApJ...804L...7Z}. 

Some of the strongest PAH emission spectra are observed around \HII\ regions, where the region immediately beyond the ionized region (often called a photodissociation region or PDR; e.g., \citealt{1985ApJ...291..722T}) has the right conditions for the existence and excitation of the molecules. Many studies have examined the appearance of the PAH bands in these environments, however there have only been a few which have done this in a spatially resolved way (e.g., \citealt{2009ApJ...706L.160B, 2012ApJ...753..168B, 2013ApJ...771...72S, 2014ApJ...791...99S}). The ability to examine PAH emission in a spatially resolved observation has become very important in developing the PAH model because the emission from the PAH molecules is sensitive to the physical conditions in which they reside and the variations seen therefore track changes in the local conditions.

In this paper, which will be the first of a series investigating aspects of the data presented here in detail, we describe the acquisition, reduction and measurement of \textit{Spitzer}/IRS observations for a sample of observed \HII\ regions, as well as several famous RNe for comparison. Our analysis will focus on their MIR emission characteristics, particularly the aromatic bands at 6.2, ``7.7", 8.6, 11.2 and 12.7 \micron. The remainder of this paper will be organized as follows: in Section~\ref{sec:data} we will discuss all aspects of our data, from the targets themselves, to the observational properties and data reduction. Section~\ref{sec:ma} contains a detailed description of all our measurement and analysis techniques, and Section~\ref{sec:res} describes our results. Finally, Sections~\ref{sec:disc} and \ref{sec:cs} provide some brief discussion of our results and then a summary of our conclusions.

\section{Data}\label{sec:data}

\subsection{Target Selection and Observations}

\begin{figure*}
    \centering
    \includegraphics[width=17cm]{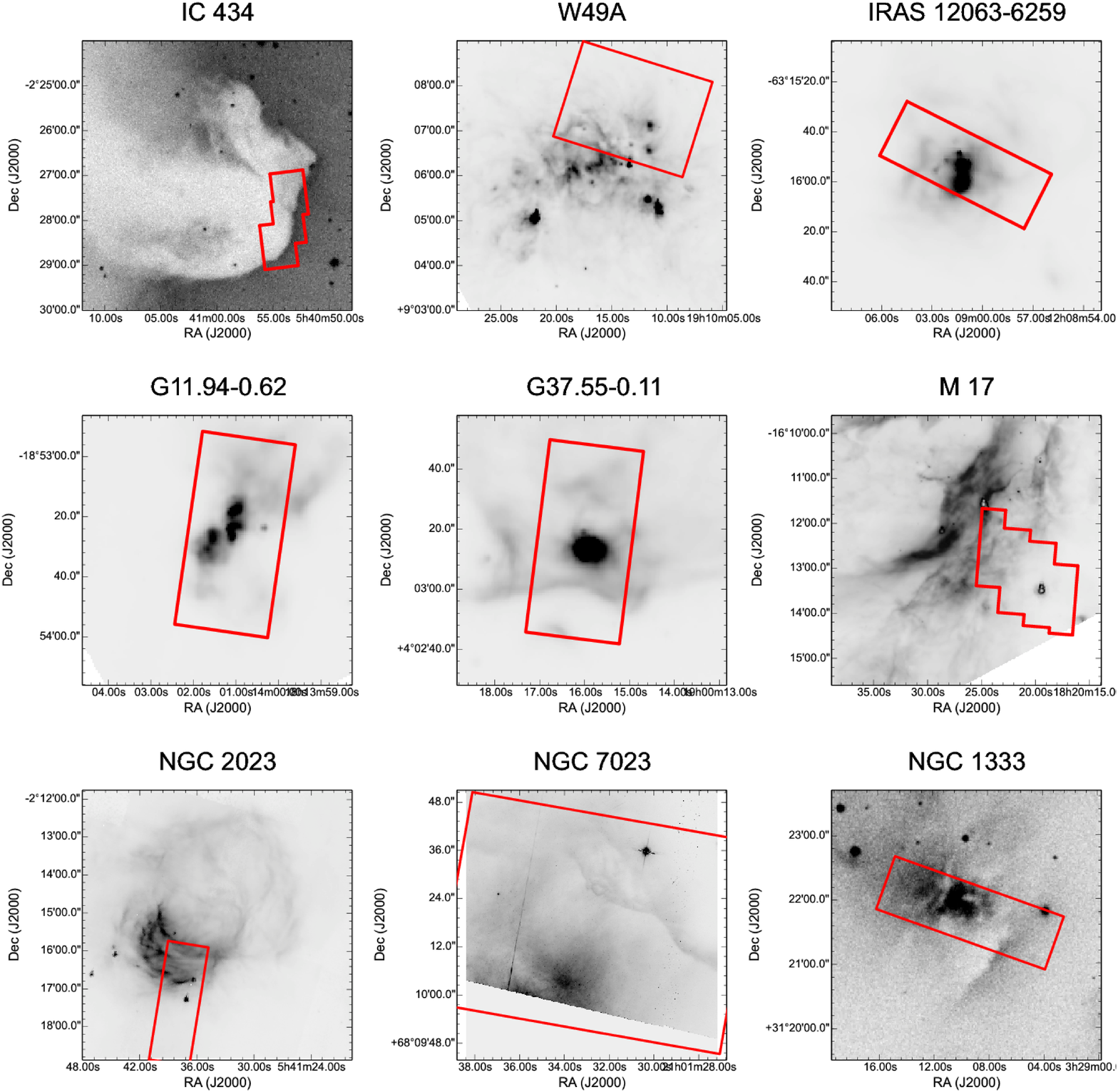}
    \caption{SL apertures. Top row: Horsehead / IC~434, W49A, IRAS~12063-6259; middle row: G11.94-0.62, G37.55-0.11, M17; bottom row: NGC~2023, NGC~7023, NGC~1333. For each image north is up and east is to the left. The images show the apertures of our final SL cubes superimposed upon IRAC 8 \micron\ band imagery from the Glimpse survey \citep{2003PASP..115..953B,2009PASP..121..213C} with the exception of the Horsehead / IC~434 field; NGC~1333 for which we have used Supercosmos sky survey R band imagery \citep{2001MNRAS.326.1279H}; and the NGC 7023 field for which we have used HST imagery. }
    \label{fig:pointings}
\end{figure*}

We have selected the mapped \HII\ regions from the \textit{Spitzer} archive, along with a subset of well-known RNe. Our objects span a variety of classes and physical properties, from the ultra-compact IRAS~12063-6259 to the giant star formation complex W49A (summarized in Table~\ref{tables:objects}). For each object we list its distance, the spectral type of its exciting star and its size in both angular and physical units. In addition we show the physical size of a \textit{Spitzer}/IRS pixel, as the wide spread of distances evident in Table~\ref{tables:objects} corresponds to large changes (a factor of 30) in the physical area represented by the 1.8$^{\Box}$\arcsec\ \textit{Spitzer} IRS pixels. Each of the nebulae were observed using the spectral mapping mode of the IRS instrument and its SL module, which covers the 5--15 \micron\ region at a spectral resolution of $\sim$ 128. All of the pointing and exposure time information for each map is presented in Table~\ref{tables:observations}. We also show the extent of each cube superimposed on optical/MIR imagery in Figure~\ref{fig:pointings}. Rather than discuss each source in detail here, we refer to final column of Table~\ref{tables:observations}, where we list the papers which have previously examined these specific spectral cubes and present discussion of the various objects. As the G37.55-0.11, G11.94-0.62, M~17 and NGC~1333 observations have not hitherto been presented, we give a brief overview of these sources here.

\paragraph{G11.94-0.62 and G37.55-0.11} These sources, which are also known as IRAS~18110-1854 and IRAS~18577+0358 respectively, are both ultra-compact \HII\ regions (UC-\HII) which have been extensively studied in the radio. The radio observations provide us with measures of the distance, radius and implied spectral type of the dominant exciting star (e.g., Table~\ref{tables:objects}). Neither has been spectroscopically investigated in the MIR, although both objects have been imaged in the MIR. \citet{2003MNRAS.343..143C} presented observations of both sources using MSX observations at a scale of around $\sim$20\arcsec, which revealed that the structure of the emission for G11.94 did not mirror the radio emission (which traces the ionized region), while for G37.55 the core-halo radio morphology was not contradicted by the `extended' appearance in the MIR. For G11.94 this conclusion was confirmed by \citet{2003ApJ...598.1127D} who presented much higher resolution images taken at ground based facilities. Their 20.8 \micron\ image clearly shows that there is a very strong peak in the emission close to the south edge of the cometary halo observed in the radio (e.g., \citealt{1989ApJS...69..831W}) with some extended emission to the north going well beyond the radio emission. In addition there appears to be a secondary peak in the MIR emission on the northern edge of the radio halo. The Spitzer imagery presented in Figure~\ref{fig:pointings} agrees with the MSX findings in that it shows distinct clumps of emission, in tension with the \citet{1989ApJS...69..831W} categorization of the region as `cometary'. On the other hand the image we present of G37.55-0.11 matches the expected core-halo morphology very well.

\paragraph{NGC~1333} NGC~1333 is a large, active star forming region in the Perseus molecular cloud (see \citealt{2008hsf1.book..346W} for a recent review). While the designation NGC~1333 has come to be used for the entire region, it originally refers to the bright reflection nebula in its core. The RNe is illuminated by the B8 type star BD~+30$^\circ$549 which lies around 3.5 \arcmin ($\sim$0.23 pc) to the north east of the RNe and an early B star known as ``SVS3"\footnotemark\ \citep{1976AJ.....81..314S}. SVS3 has been observed in the radio and MIR, and found to have a very compact \HII\ region consistent with the central star being of spectral type B4-6 \citep{1984ApJ...278..156H, 1986ApJ...303..683S}. The region around this source has also been spectroscopically observed in the MIR, with several studies in the 90's using such observations to make the first claims of the presence of ionized PAH molecules in the ISM (e.g., \citealt{1996ApJ...460L.119J, 1999ApJ...513L..65S}). Both of these studies used ground based observations of the 8--13 \micron\ region and so were limited in aperture. Despite that limitation, both studies found that the 8.6 \micron\ band peaked in the center of the region (at around the location of SVS3) while the 11.2 \micron\ band peaked around 10\arcsec\ to the south (corresponding to a physical distance of about 0.64 pc). 

\footnotetext{This source has been given a variety of different designations (e.g., LZK 12; IRAS~03260+3111), we shall refer to it as SVS3 as this appears to be the common usage for MIR observations. }

\begin{table*}
\caption{\label{tables:objects}Target Properties}
\begin{center}
\begin{tabular}{l c c c c c c c c}
\hline\hline
Object            & Distance & Exciting Star & Angular Size$^a$         & Physical Size       & Pixel Size & References \\
                  &  [kpc]   & Spectral Type & \arcmin $\times$ \arcmin & [pc] $\times$ [pc]  & [pc] &\\ 
\hline
\\
\textsl{H~\textsc{ii} Regions:}\\
\hline
W49A              & 11.4     & O5            & 7 $\times$ 5             & 23 $\times$ 17   & 0.20  & 1, 2 \\
IRAS~12063-6259   & 10.9     & $>$O5.5$^b$  & 1 $\times$ 1             & 3.2 $\times$ 3.2 & 0.19  & 3 \\
G37.55-0.11       & 10.1     & O6            & 2 $\times$ 1.5           & 5.9 $\times$ 4.4 & 0.18  & 4, 5 \\
G11.94-0.62       & 5.2      & O7.5          & 1.1 $\times$ 0.7         & 1.7 $\times$ 1.1 & 0.09  & 6\\
M~17              & 1.6      & O4            & 10 $\times$ 10           & 4.7 $\times$ 4.7 & 0.03  & 7, 8\\
Horsehead         & 0.42     & O9            & 4 $\times$ 1.5           & 0.5 $\times$ 0.2 & 0.007 & 9, 10\\
\\
\textit{Reflection Nebulae:}\\
\hline
NGC 7023          & 0.43     & B2.5          & 3 $\times$ 2.2           & 0.4 $\times$ 0.3 & 0.008 & 11\\
NGC 2023          & 0.35     & B1.5          & 6 $\times$ 4.5           & 0.6 $\times$ 0.4 & 0.006 & 12, 13\\
NGC 1333          & 0.23     & B8            & 2 $\times$ 2             & 0.1 $\times$ 0.1 & 0.004 & 14, 15\\
\\
\hline\hline
\end{tabular}
\end{center}
\textit{a}: Measured using IRAC 8 \micron\ band imagery so as to be comparable to that expected in IRS cubes.\\
\textit{b}: IRAS~12063-6259 is thought to be illuminated by a cluster of young stars (e.g., \citealt{2003A&A...407..957M}), however as we do not possess individual radio fluxes here we quote the spectral type of a single star producing the same radio luminosity (using the model grids of \citealt{1996ApJ...460..914V}) as an upper limit.\\

\tablerefs{(1) \citet{1997ApJ...482..307D}; (2) \citet{1992ApJ...393..149G}; (3) \citet{1987AA...171..261C}; (4) \citet{2003ApJ...587..714W}; (5) \citet{2001ApJ...549..979K}; (6) \citet{1989ApJS...69..831W}; (7) \citet{2003ApJ...593..874T}; (8) \citet{2008ApJ...686..310H}; (9) \citet{2008AJ....135.1616S}; (10) \citet{1976AJ.....81..245E}; (11) \citet{1997AA...324L..33V}; (12) \citet{2009AA...507.1485M}; (13) \citet{2011ApJ...741...45S}; (14) \citet{1968AJ.....73..233R}; (15) \citet{2008PASJ...60...37H}.}
\bigskip
\end{table*}

\begin{table*}
\caption{\label{tables:observations}Observation Properties}
\begin{center}
\begin{tabular}{l p{1cm} l p{1.5cm} c c c c c c}
\hline\hline
Object            & IRS     & Program & AORkey   & Position Angle  & Steps       & Steps  & Step size              & Step size         & Reference \\
                  & Module  & ID      &          & [$^\circ$]      & $\parallel$ & $\bot$ & $\parallel$ [\arcsec] & $\bot$ [\arcsec] &         \\
\hline
\\
\textsl{H~\textsc{ii} Regions:}\\
\hline
W49A              & SL      & 63      & 16206848 & 84              & 5           & 100    & 40                    & 1.8              & 1\\
G37.55-0.11       & SL      & 20517   & 14799104 & -93             & 1           & 9      & --                    & 3.6              & -- \\
IRAS~12063-6259   & SL      & 20517   & 14798848 & 157             & 1           & 7      & --                    & 3.6              & 2\\
G11.94-0.62       & SL      & 20517   & 14799360 & 86              & 1           & 9      & --                    & 3.6              & -- \\
M~17              & SL      & \\
Horsehead         & SL      & 3512    & 12011520, 12017920, 12018688  & 101             & 1           & 23     & --                    & 1.8              & 9, 10\\
\\
\textit{Reflection Nebulae:}\\
\hline
NGC 2023 S        & SL      & 30295   & 17977856 & 84              & 3           & 18     & 28                    & 3.6              & 4\\
NGC 7023          & SL1, SL2     & 28      & 3871744, 3871488  & -1.2            & 2           & 15     & 7.3                   & 3.6              & 6, 7\\
NGC 1333          & SL      & 20378   & 14587648 & -107            & 5           & 111    & 54.6                  & 1.8              & --     \\
\\
\hline\hline
\end{tabular}
\end{center}
\tablecomments{The M17 and Horsehead IRS-SL AORs are combined to create the final maps. The NGC~7023 SL and SH cubes are sometimes analyzed after being stitched together (e.g., 6) as the cubes overlap to a high degree. For the other sources we have not merged the SL/SH cubes as this would involve discarding as much as 70\% of the SL cubes in some cases. }
\tablerefs{(1) \citet{2014ApJ...791...99S}; (2): \citet{2013ApJ...771...72S}; (3) \citet{2012ApJ...747...44P}; (4) Peeters et al. 2015; (5) \citet{2011A&A...532A.128R}; (6) \citet{2014ApJ...795..110B}; (7) \citet{2007AA...469..575B}; (8) Shannon et al. 2015; (9) \citet{2014A&A...563A..65O}; (10) \citet{2015A&A...576A...2O}     }

\end{table*}

\subsection{Data Reduction}

The spectral maps for each object were processed using the \textit{Spitzer} Science Center reduction pipeline for IRS/SL (generally Version 18.8, 18.7 in some instances). CUBISM \citep{2007PASP..119.1133S} was employed to clean the individual BCDs, primarily this meant masking any rogue or otherwise `bad' pixels, initially using the CUBISM functions `AutoGen Global Bad Pixels' with settings `Sigma-Trim' = 7, `MinBad-Frac' = 0.5 and `AutoGen Record Bad Pixels' with settings `Sigma-Trim' = 7 and `MinBad-Frac' = 0.75 and subsequently by manual inspection of each cube. This process was repeated for each order (SL1, SL2, SL3) of each observation for each object. In order to combine the cubes of different orders, SL2 and SL3 were resampled to the same pixel grid as SL1 using CUBISM, then the scaling factors were calculated between the three orders for the now aligned pixels by comparing the average fluxes within the overlap regions. Subsequently the SL1 and SL2 cubes were stitched together using the derived scaling factors. For each of our maps the scale factors between orders were around 10\%, with outliers being investigated and in some cases flagged as poor data. 

As the IRS/SL pixels are around half the size of the beam, the final cubes were created by averaging 2$\times$2 pixel apertures, resulting in overlapping extraction apertures in which one in every four pixels is independent. This was performed by taking the average of the next three adjacent pixels, i.e., $F_1(x,y,\lambda) = 0.25 \times (F_0(x,y,\lambda) + F_0(x+1,y,\lambda) + F_0(x,y+1,\lambda) + F_0(x+1,y+1,\lambda))$, where $F_1(x,y,\lambda)$ is the resulting cube and $F_0(x,y,\lambda)$ is the observed cube. This process produces a new cube $F_1(x,y,\lambda)$ which has the same pixel scale as the old cube, however each pixel represents a local beam average. Subsequent analysis masked out the non independent pixels. While several of the maps were not Nyquist sampled, we adopted the same procedure such that we produced a uniform sample of spectral maps.

\section{Measurement \& Analysis}\label{sec:ma}

The analysis of the spectra took place in several stages. Firstly, we measured and corrected for extinction (as will be described in Section~\ref{sec:ex}). Secondly, we defined the continua and aromatic emission plateaus using spline fits. Thirdly, we fit the continuum subtracted emission features in a variety of ways, from direct integration for some PAH bands, to Gaussian fits for the emission lines. In general this combination of techniques is referred to as the `Spline' method (e.g., \citealt{2000A&A...357.1013V, 2001A&A...370.1030H,2002A&A...390.1089P, 2004ApJ...611..928V, 2010A&A...511A..32B, 2013ApJ...771...72S, 2014ApJ...795..110B, 2014ApJ...791...99S, 2015ApJ...811..153S}). Other methods of fitting the PAH spectra involve fitting an array of Drude profiles to model the PAH emission (e.g., PAHFIT; \citealt{2007ApJ...656..770S}), or using templates based on principle component analysis (e.g., PAHTAT; \citealt{2012A&A...542A..69P}). It has been shown though that the trends present in the PAH band intensity measurements do not depend on the measuring technique employed (e.g., \citealt{2008ApJ...679..310G,2012ApJ...747...44P}), and so we are confident that our use of the Spline method does not introduce any significant biases.

Before processing the data for each cube though, regions of poor data or contamination from unwanted sources were masked out. Primarily this meant removing the pixels associated with young stellar objects (YSO's) in M~17 and NGC~2023 (visible as bright sources towards the middle of the NGC~2023 south cube and in the middle of the M~17 cube). In addition we removed sections of cubes which had fallen `off-source', i.e., pixels which did not contain significant flux -- such as the southern half of the NGC~2023 south cube, and the northern and southern 20\% of the NGC~1333 cube.

\subsection{Continua and Extinction}\label{sec:ex}
In general, molecular clouds and \HII\ regions contain copious amounts of dust, which combine to create both continuous extinction from the UV to the mid-infrared (although this is generally a small effect in the MIR) as well as strong absorption features associated with dust grains of specific compositions. The most important of these, from the point of view of measuring PAH band fluxes, is the silicate feature at 9.8 \micron. This feature has significant width which overlaps two important PAH bands at 8.6 and 11.2 \micron. Typically the 11.2 \micron\ band is used to normalize measurements of the other bands, so its obscuration can strongly effect the interpretation of the PAH spectrum. 

In order to correct our band measurements we adopt the `modified Spoon method' from \citet{2013ApJ...771...72S, 2014ApJ...791...99S} which is an updated version of the method used by \citet{2007ApJ...654L..49S} which includes the silicate absorption profile given by \citet{2006ApJ...637..774C}. This procedure relies on the 5-15 \micron\ continuum displaying a strong rising power law towards the red, which is appropriate for the \HII\ regions. However, the RNe do not generally display spectra with rising power law continua, instead they seem to have roughly linear continua in physical (i.e. W m$^{-2}$ \micron$^{-1}$ vs. \micron) units. Therefore for the RNe we have used a modified version of our extinction measurement routine in which we assume a linear continuum between 5 and 15 \micron. This assumption is based on the observational results of samples of PDRs (e.g., \citealt{2007AA...469..575B, 2008AA...491..797C}). 
\subsubsection{HII Region Extinction}

\begin{figure*}
	\begin{center}
		\includegraphics[width=17cm]{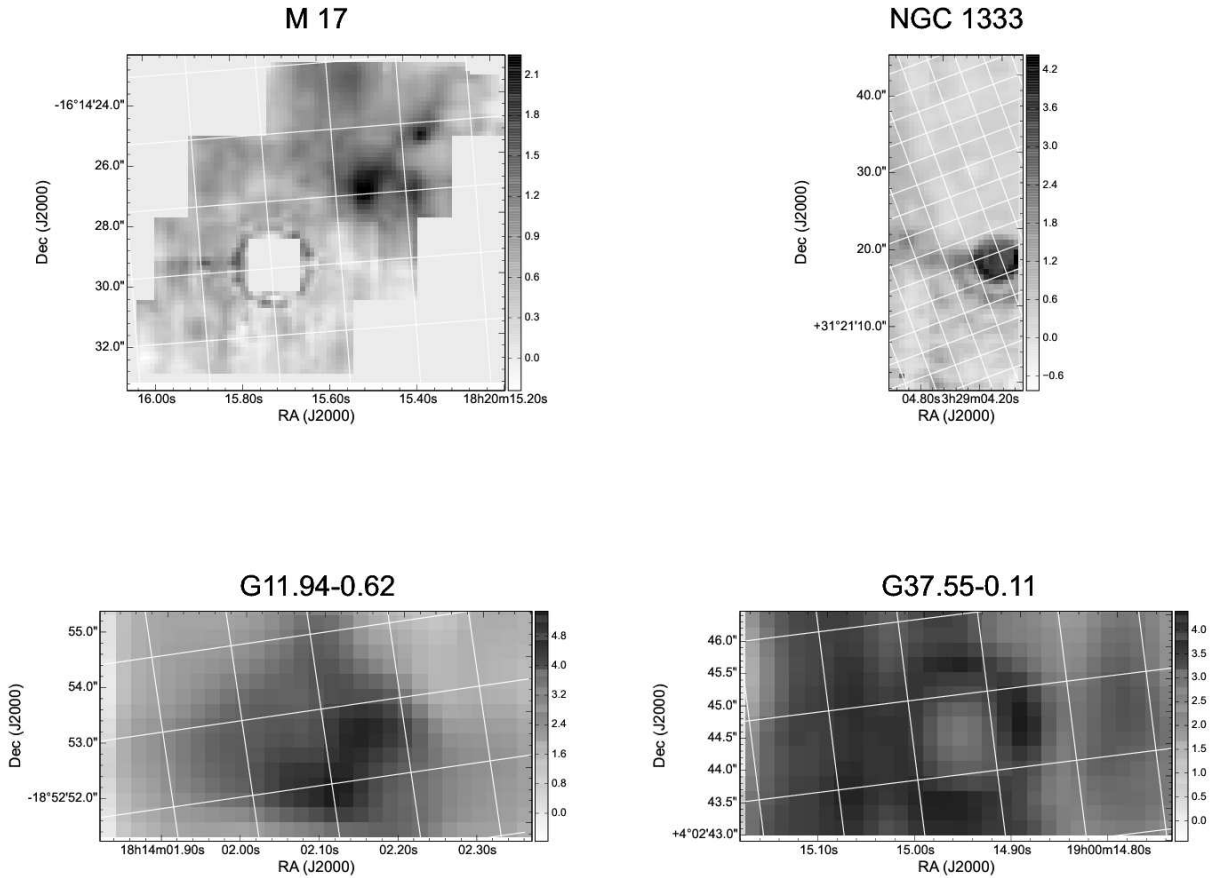}
	\end{center}	
	\caption{Extinction maps in units of A$_K$. Top: M~17, NGC~1333; bottom: G11.94-0.62, G37.55-0.11. The grid overlaid on each image shows constant RA and Dec, with the approximately vertical lines being aligned north-south.}
	\label{fig:ext_maps}
\end{figure*}

The extinction maps for the \HII\ regions presented, i.e. G11.94-0.62, G37.55-0.11 and M17, are shown in Figure~\ref{fig:ext_maps} (along with the RNe, which will be discussed in the following section). The Horsehead PDR region was found to have essentially zero extinction measurable via the silicate feature.

\paragraph{W49A and IRAS 12063-6259} The W49A and IRAS~12063-6259 extinction maps have been shown and discussed at length by \citet{2013ApJ...771...72S} and \citet{2014ApJ...791...99S}, respectively. In general each of the maps was found to be consistent with prior measurements of stars on the line of sight or the nebulosity itself.

\paragraph{M~17} The M~17 extinction map displays reasonably consistent extinction levels throughout, ranging from A$_K$ of 0.5 to a maximum of 2 in a region close to the main PDR in the northern half of the map. The north eastern corner of the map, which emits the majority of the PAH flux has an average A$_K$ of around 1. In the southern half of the map we have masked out a protostar that dominated our extinction measurements. 

\begin{figure}
	\begin{center}
		\includegraphics[width=7cm]{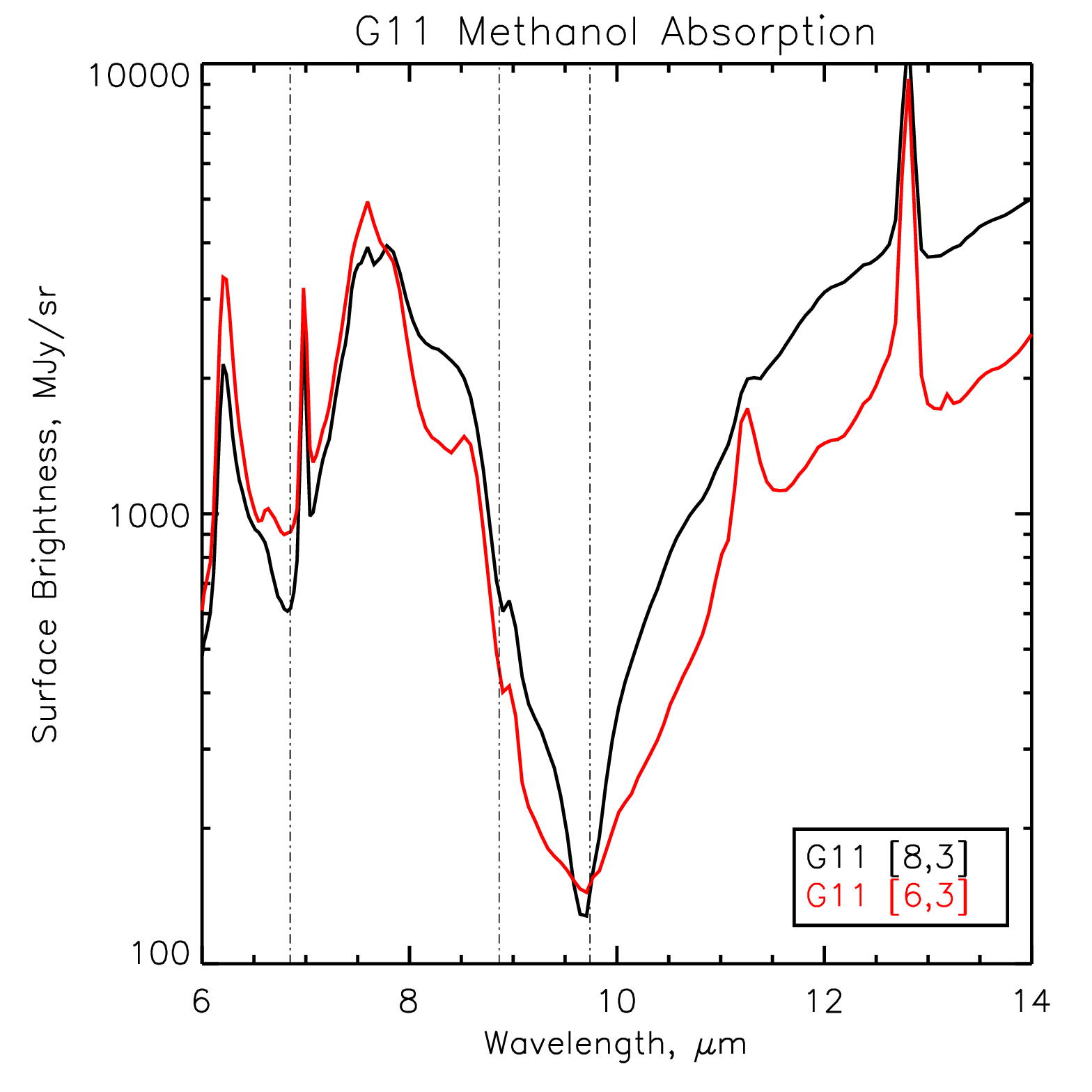}
	\end{center}	
	\caption{Probable CH$_3$OH absorption in the G11.94-0.62 spectra. The red spectrum is a `typical' \HII\ region spectrum with heavy silicate absorption and the black spectrum is similar except for prominent absorption bands at 6.8 and 9.7 \micron, which are much narrower than the silicate absorption. The vertical dot-dashed lines mark the locations of CH$_3$OH absorption bands given by \citet{1992ApJ...399L..79S} and the pixels from which the spectra were drawn given in the key.}
	\label{fig:g11_meth}
\end{figure}

\paragraph{G11.94-0.62} Extinction in G11.94-0.62 is strongly peaked around the core of the \HII\ region (at around A$_K$ = 4), which drops off very strongly to less than A$_K$ = 2 within a few pixels. The center of the map is affected by CH$_3$OH absorption at 9.8 \micron, which is much narrower than the silicate absorption and clearly evident in the spectra (e.g., \citealt{1992ApJ...399L..79S}; \citealt{2011A&A...526A.152V}; also Figure~\ref{fig:g11_meth}). In order to measure the silicate absorption for G11.94-0.62, we modified the general method described in our previous papers (e.g., \citealt{2013ApJ...771...72S, 2014ApJ...791...99S}) such that the wavelength of comparison between the observations and the interpolated power law spectra was outside the methanol band, in this case at 10.2 \micron\ rather than 9.8 \micron. The spectra strongly affected by methanol absorption were not used in the remaining analysis, as the methanol absorption created difficulties in defining an effective spline continuum for isolating the PAH bands.

\paragraph{G37.55-0.11} An almost complete ring of silicate absorption is visible around the center of G37.55-0.11. This could represent the projection of a spherical dust shell, as the central region still retains a high extinction level of around A$_K$ = 2 compared to the silicate absorption ring, which peaks at around 3.5

\subsubsection{RNe Extinction}

\paragraph{NGC 2023} Extinction values found for NGC~2023 are very low (of the order of A$_V$ $\sim$ 1; A$_K$ $\sim$ 0.1), which is slightly lower than the average from previous measurements (e.g., \citealt{2011ApJ...741...45S} and references therein). As the value of A$_K$ recovered was very low, we did not deredden the NGC~2023 cube, as doing so would have introduced an additional source of uncertainty while only affecting the fluxes very minimally. \citet{2012A&A...542A..69P} used the PAHTAT tool and found values of A$_K$, which were systematically slightly higher than both our measurements and those discussed by \citet{2011ApJ...741...45S}. 

\paragraph{NGC 7023} We find A$_V$ values of up to around 20 in the north-west corner of the map behind the PDR front in agreement with that found by \citet{2013ApJ...769..117B}, who used a method similar to that employed by PAHFIT, albeit using a different extinction law. In contrast, the extinction map presented by \citet{2012A&A...542A..69P} finds values of up to A$_V$ $\sim$ 40, a factor of two more than our measurement, albeit with the same general distribution. We also ran PAHFIT on each pixel of the cube and recovered the same pattern of extinction, except with a higher still peak value of around A$_V$ = 60. We dereddened the NGC~7023 cube using our extinction measurements.

\paragraph{NGC 1333} The NGC~1333 extinction map shows that the core of NGC~1333 (the faint ring in the center of the map) suffers from negligible extinction (A$_K$ $<$ 0.5 ), as we might expect given that it is a nearby RNe. However, the western side of the map is dominated by a filament of the molecular cloud material, which obscures part of the field with A$_K$ $>$ 3. In particular this ridge is inhabited by the protostar IRAS~03260+3111E -- which has the strongest silicate absorption in the spectral cube. Spectra associated with this object were excluded from further analysis. For NGC~1333 we dereddened the cube using our measurements.

\subsection{Band Measurements}
\begin{figure*}
	\begin{center}
	\includegraphics[width=18cm]{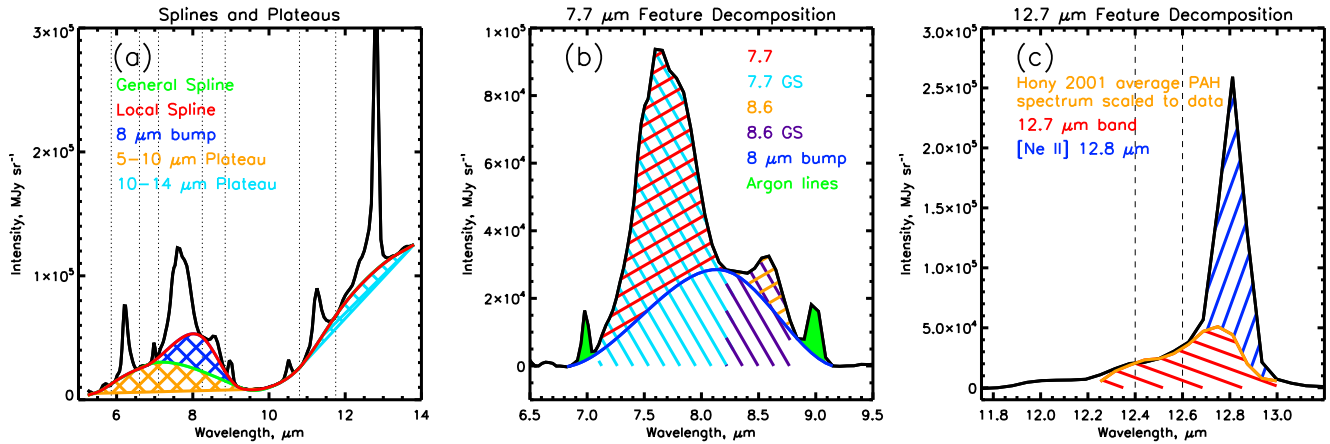}
	\end{center}
	\caption{Examples of the spline spectral decomposition method. (a): The primary adopted splines are shown in red (local) and green (general), along with the plateaus defined by the difference between these splines (the 8 \micron\ bump) and between the general spline and straight lines connecting the continuum under the primary features (5-10 \micron\ and 10-14 \micron\ plateaus). (b): Traditional decompositions of the `7.7' feature using the spline method, both including (`7.7', `8.6') and excluding (`7.7~GS', `8.6~GS') the flux of the 8 \micron\ bump. (c) Decomposition of the $\sim$12.7 \micron\ emission into the 12.7 \micron\ PAH band and the [Ne~{\sc ii}] 12.8 \micron\ line using the average PAH spectrum given by \citet{2001A&A...370.1030H}.}
\bigskip
	\label{fig:decomp}
\end{figure*}

The specifics of our implementation of the spline method are identical to those presented by \citet{2014ApJ...791...99S}. In short, this process involves computing a spline fit to a set of continuum points at wavelengths $\lambda$ = 5.37, 5.46, 5.86, 6.58, 6.92, 8.28, 9.15, 9.40, 9.64, 10.14, 10.33, 10.76, 11.82, 13.18, 13.49, 14.43, 14.74 \micron, and subtracting this spline from the measured data in order to create a continuum subtracted spectrum. In Figure~\ref{fig:decomp} we show an example of the spline-method applied to a W49A spectrum, as well as examples of the techniques used to decompose the 7.7 \micron\ complex and the blend between the 12.7 \micron\ PAH band and the [Ne~{\sc ii}] 12.8 \micron\ atomic line. 

In general two splines were defined for each spectrum, which differ from each other by the inclusion of a spline point at 8.28 \micron\ (between the peaks of the 7.7 and 8.6 \micron\ PAH bands). The first plateau, which we refer to as the 8 \micron\ bump, is simply the difference of the two splines. The remaining plateaus that underlie the 5--10 \micron\ and 10--14 \micron\ regions are defined by drawing a straight line underneath the features and integrating all of the flux under the general spline. This process is displayed graphically in Figure~\ref{fig:decomp}. Measurement of the narrower emission features proceeds using continuum subtracted spectra.

Measurement of the PAH bands then proceeds in three distinct ways. The fluxes of the strong bands at 6.2, `7.7', 8.6, and 11.2 \micron\ are found by direct integration. For the 6.2 and 11.2 \micron\ bands these measurements are supplemented with Gaussian profile fits to determine the strengths of the weak 6.0 and 11.0 \micron\ bands with which each are blended. The final 6.2 and 11.2 band fluxes are then found by subtracting the weaker bands from the total flux. While we have measured the 6.0 \micron\ PAH band, for the sake of brevity we do not discuss it, or the other weak PAH bands at 5.7, 12.0, 13.5 and 14.0 \micron.

For the `7.7' complex the flux was directly integrated to produce measurements of the 7.7 and 8.6 \micron\ bands. However, as here is some disagreement in the literature as to how the plateau underlying the `7.7' \micron\ complex is related to the PAH bands, this measurement was repeated for spectra which had been continuum subtracted using both the local and general spline -- producing two measurements of both the 7.7 and 8.6 \micron\ bands, i.e. with and without a spline point at 8.3 \micron\footnotemark.

\footnotetext{Which we refer to as the 7.7 \micron\ band measurement and the 7.7GS \micron\ band measurement.}

For the \HII\ regions\footnotemark the blend between the 12.7 \micron\ PAH band and 12.8 \micron\ [Ne~{\sc ii}] line is decomposed using the method proposed by \citet{2014ApJ...791...99S} (and subsequently used by \citealt{2015ApJ...811..153S}) in which the 12.7 \micron\ emission is assumed to have the same spectral profile as the average PAH emission spectrum given by \citet{2001A&A...370.1030H}. The Hony profile is scaled to the continuum subtracted emission in a small window between 12.4 and 12.6 \micron\ where the emission is assumed to be dominated by the PAH band (shown in Figure~\ref{fig:decomp} (c)). The PAH flux can then be recovered by integrating under the feature profile and the 12.8 \micron\ atomic line can then be measured using a simple Gaussian profile fit after the PAH emission profile has been subtracted. While it would have been possible to directly integrate the remaining flux to obtain the 12.8 \micron\ line flux, we chose to use a Gaussian profile fit in order to ensure that the the method of subtracting a scaled 12.7 \micron\ PAH emission band shape does not affect the shape of the 12.8 \micron\ line. The possible systematics affecting this measurement (i.e. the presence of very strong [Ne~{\sc ii}] emission) are discussed in detail by \citet{2015ApJ...811..153S}. For the RNe, the 12.7 \micron\ band flux was calculated by directly integrating the complex and subtracting the flux of the 12.3 \micron\ H$_2$ line which was fitted using a Gaussian profile fit superposed on a rising polynomial continuum. 

\footnotetext{The atomic emission line is not present for the RNe, and hence the 12.7 \micron\ feature for the RNe could be measured using direct integration.}

The other atomic lines, where present, are all measured using Gaussian profile fits with either flat or polynomial continua depending on the line. For example, the 8.99 \micron\ [Ar~{\sc iii}] line sits on the rising wing of the 8.6 \micron\ PAH band and so requires a second order polynomial continuum.

In all cases uncertainties were computed by comparing the integrated feature flux and the rms noise of featureless areas of continuum between 9.3 and 9.5 \micron, 13.3 and 13.5 \micron\ and 13.7, 13.9 \micron\ respectively. We adopted this approach such that each flux measurement has an associated signal to noise ratio measurement calculated in a uniform way across the sample, rather than using, for example, the fit-parameter uncertainties for bands fit with Gaussian profiles. In all subsequent discussion the uncertainties quoted or shown for a band measurement were calculated in this way.

\section{Results}\label{sec:res}

\subsection{PAH Ratio Correlations}

\begin{figure*}
	\begin{center}
	\includegraphics[width=17cm]{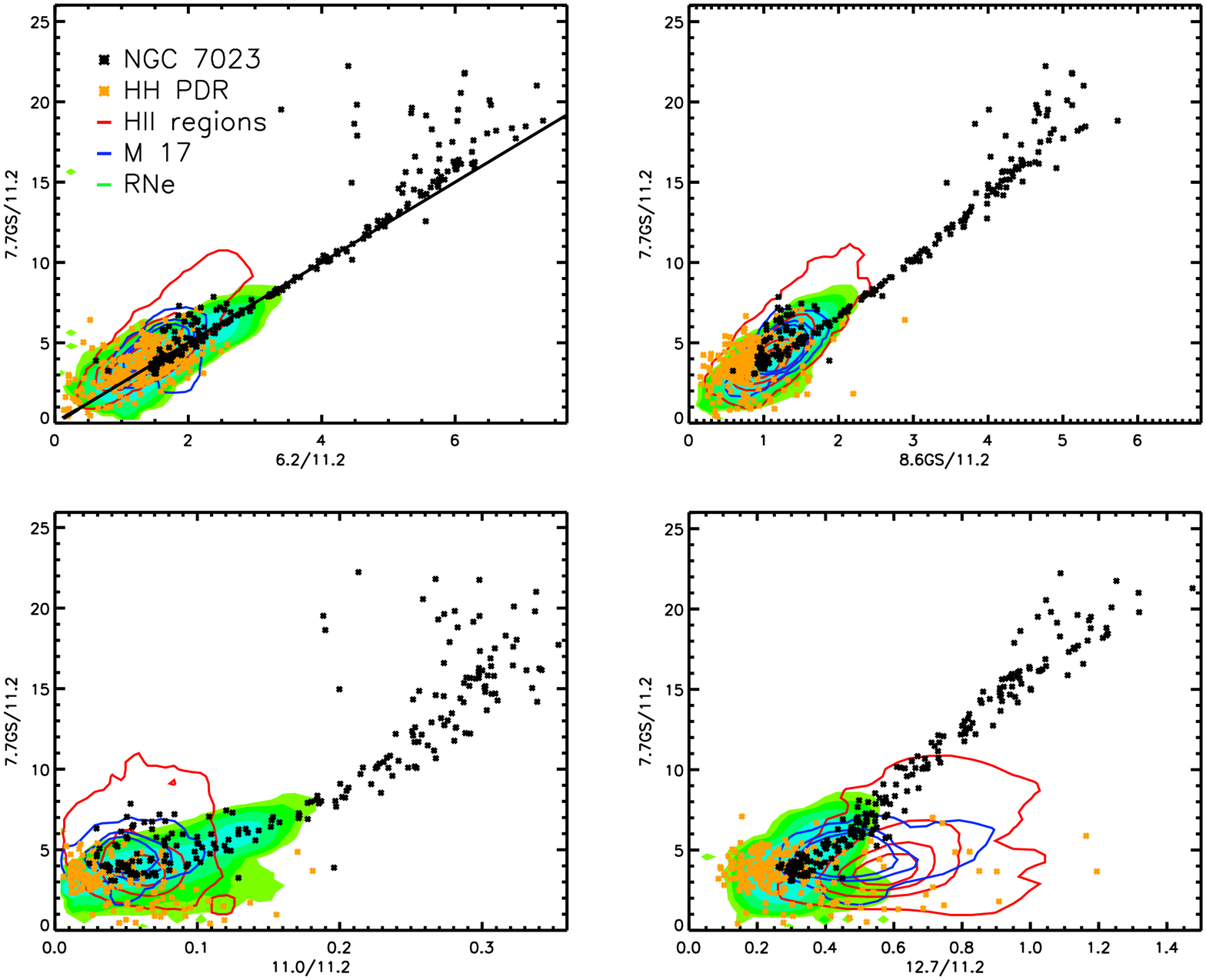}
	\end{center}
	\caption{PAH band ratio correlations, where for each panel we have only included points where the S/N of our measurement of the ratio exceeds three. For the 6.2 vs. 7.7 correlation we have included a straight line generated by assuming a constant 6.2 vs. 7.7 band ratio and variable 11.2 band strength. Combined \HII\ region points - red contours; NGC~1333 and 2023 - green to blue filled contours; NGC~7023 - black points; Horsehead - orange points; M17 - blue contour, contour levels given in Table~\ref{tables:contlevs}.}
	\label{fig:rat_corr}
\end{figure*}

The standard correlations between PAH ratios, obtained by `normalizing' each PAH flux by dividing by the flux of another PAH band in the same spectrum, are presented in Figure~\ref{fig:rat_corr}, where we have chosen to use the 11.2 \micron\ band to normalize the data. This procedure is performed to cancel out the effects of, for example, PAH abundance and distance. In Figure~\ref{fig:rat_corr} (and the following Figures~\ref{fig:rat_corr_plat} and \ref{fig:lm_corr_all} we have chosen to present the majority of the data as contours for the sake of clarity. This was achieved by binning the data for each object into 50 bins in both $x$ and $y$ directions across the range of values present. Subsequently these binned data were slightly smoothed using a 2 pixel boxcar average routine and plotted as contours. The contour levels are given in Table~\ref{tables:contlevs} and were adjusted such that they only exclude isolated outliers. In Figure~\ref{fig:rat_corr} we compare each MIR emission component in turn to 7.7GS / 11.2. For this figure we have included only the points where the signal-to-noise ratio for the combined bands is greater than three. In Figure~\ref{fig:rat_corr} we have chosen to represent the data by grouping the points of several of the sources which display the same overall trends. The rationale for these groupings is clearer in the flux versus flux plots presented in the subsequent Section, however the overall behavior within each group is also consistent in the traditional correlation plots. The first group represents the \HII\ regions and contains W49A, IRAS~12063-6259, G11.94-0.62 and G37.55-0.11. The second major group represents the RNe and contains NGC~1333 and NGC~2023. Finally, there are three sources which resist classification into these groups, the RN NGC~7023 and the \HII\ region PDRs of M~17 and the Horsehead nebula. 

\begin{table}
\caption{\label{tables:contlevs}Contour Levels}
\begin{center}
\begin{tabular}{c c c}
\hline\hline
Object			& Figures~\ref{fig:rat_corr} \& \ref{fig:rat_corr_plat} & Figures~\ref{fig:lm_corr_all} \& \ref{fig:curveHII}\\
			& [points/bin]					& [points/bin]\\		
\hline
\HII\ Regions		& 0.5, 10, 30, 50				& 0.25, 5, 10  \\
M~17			& 0.5, 5, 10					& 0.25, 2, 5, 7 \\
NGC 2023S, NGC 1333     & 0.5, 1, 2, 4					& 0.25, 0.5, 1, 2, 4 \\
\hline\hline
\end{tabular}
\end{center}

\end{table}

\paragraph{7.7 vs. 6.2} For the most well-known correlation, between the 7.7 and 6.2 bands, the expected linear relationship is recovered in each case (albeit with a slightly different gradient for the \HII\ regions sample). For the NGC~7023 points in this plot we have shown that the majority of the points can be well explained by assuming that the 7.7 vs. 6.2 \micron\ band intensity ratio is essentially constant and varying the 11.2 \micron\ band intensity (indicated by the black line). Each of the other groups shown are also consistent with such a straight line, albeit with slight variations in the average 7.7 vs. 6.2 \micron\ band intensity ratio.

\paragraph{7.7 vs. 8.6 and 11.0} For these correlations there are some splits by object type. For example for the 7.7 vs. 8.6 correlation the RNe points have significantly better correlations than the \HII\ regions. M~17 and the Horsehead PDR in particular display almost no correlation, while the grouped \HII\ regions display smaller variation than the RNe group (NGC~1333, 2023) by around a factor of two. The latter is repeated by the 7.7 vs. 11.0 band correlation plot, where now the grouped \HII\ regions also show no correlation and the RNe display strong, linear correlations. This may be a beam dilution effect whereby the range of variation in the \HII\ region data is squashed because the measurement represents the average over a much larger physical area because the distance to the \HII\ regions is much greater than our other sources (see Table~\ref{tables:objects}). Interestingly, both of these correlations display some curvature in the correlation produced by the NGC~7023 points -- indicating that either a) our decomposition is not totally isolating the 8.6 or 11.0 \micron\ bands from the surrounding emission (i.e. both are systematically low creating a slight upward curve on the plots) or, b) that the 8.6 and 11.0 \micron\ bands do not follow the 7.7 \micron\ band as well as the 6.2 \micron\ band.

\paragraph{7.7 vs. 12.7} For the final PAH band considered, at 12.7 \micron, we see variations in behavior between the different groups. There is evidence of at least two different gradients, one of which is defined by the RNe (plus some of the W49A points) and another displayed best by M~17. Curiously, the 12.7 band appears to be much weaker for the Horsehead PDR, with most of the points having a 12.7/11.2 value of around 0.2, which is well below the average of all the points (0.4--0.5). Interestingly, for both the 8.6 and 12.7 vs. 7.7 correlation plots there is also a significant number of points for which there is variation in the 12.7/11.2 and 8.6/11.2 band ratios without any progression in the 7.7/11.2 ratio, in fact, the 7.7/11.2 band ratio seems to be about 1.5 in both cases, with the 12.7/11.2 ratio varying from 0.2 to 0.9 and the 11.0/11.2 ratio varying from 0.01 to 0.18. For the 11.0 ratio, this is exclusive to the RNe sample\footnotemark, while for 12.7 this trend is also seen for the \HII\ regions sample. 

\footnotetext{Both NGC~1333 and NGC~2023 show this trend when examined separately.}

\begin{figure*}
	\begin{center}
	\includegraphics[width=17cm]{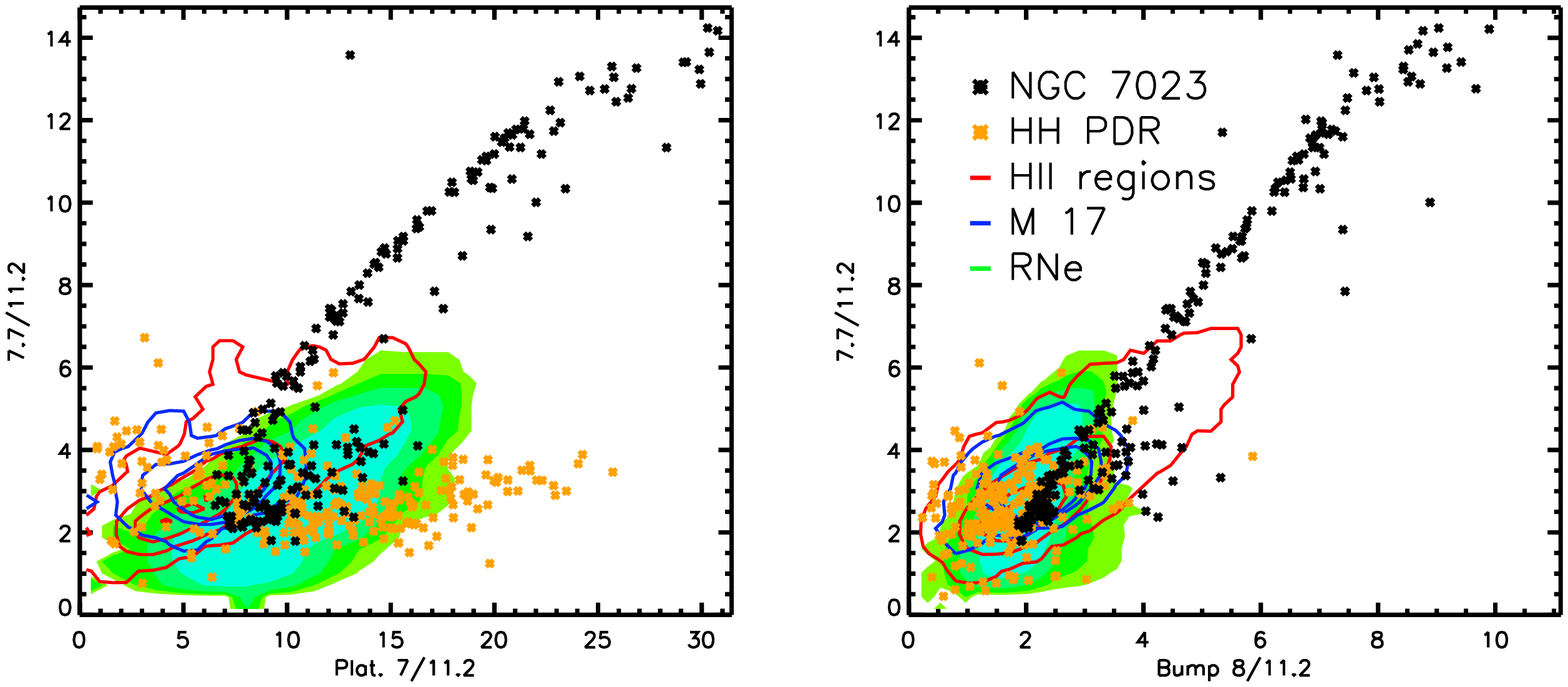}
	\end{center}
	\caption{Plateau versus PAH band ratio correlations, symbols and contours as in Figure~\ref{fig:rat_corr}, contour levels given in Table~\ref{tables:contlevs}.  }
	\label{fig:rat_corr_plat}
\end{figure*}

\paragraph{Plateaus} In Figure~\ref{fig:rat_corr_plat} we show the behavior of the 5--10 \micron\ plateau features with respect to the 7.7 \micron\ PAH band\footnotemark. Here we see some consistency between object-types with the \HII\ regions sample, M~17, the RNe sample and NGC~7023 all showing a similar trend for the 7 \micron\ plateau vs. the 7.7 band. There appears to be a difference in gradients between the RNe sample and the \HII\ regions (including M~17) though. Interestingly the Horsehead PDR points appear to support no correlation between the plateaus and the 7.7 \micron\ PAH flux. A subset of the NGC~7023 points coincides with the RNe in the 5--10 \micron\ plateau vs. 7.7 plot. The remainder of the points originate in the region of low 11.2 \micron\ band fluxes, and hence are being drawn out to the top-right of the diagram.

\footnotetext{Here we take the 7.7 \micron\ band measurements that use the local (LS) spline, as the 7.7 \micron\ band measurements using the global spline (GS) includes some of the flux which we also measure as the 8 \micron\ bump - see Figure~\ref{fig:decomp}.}

\subsection{PAH Flux Correlations}\label{sec:pahflux}

\begin{figure*}
	\begin{center}
	\includegraphics[width=17cm]{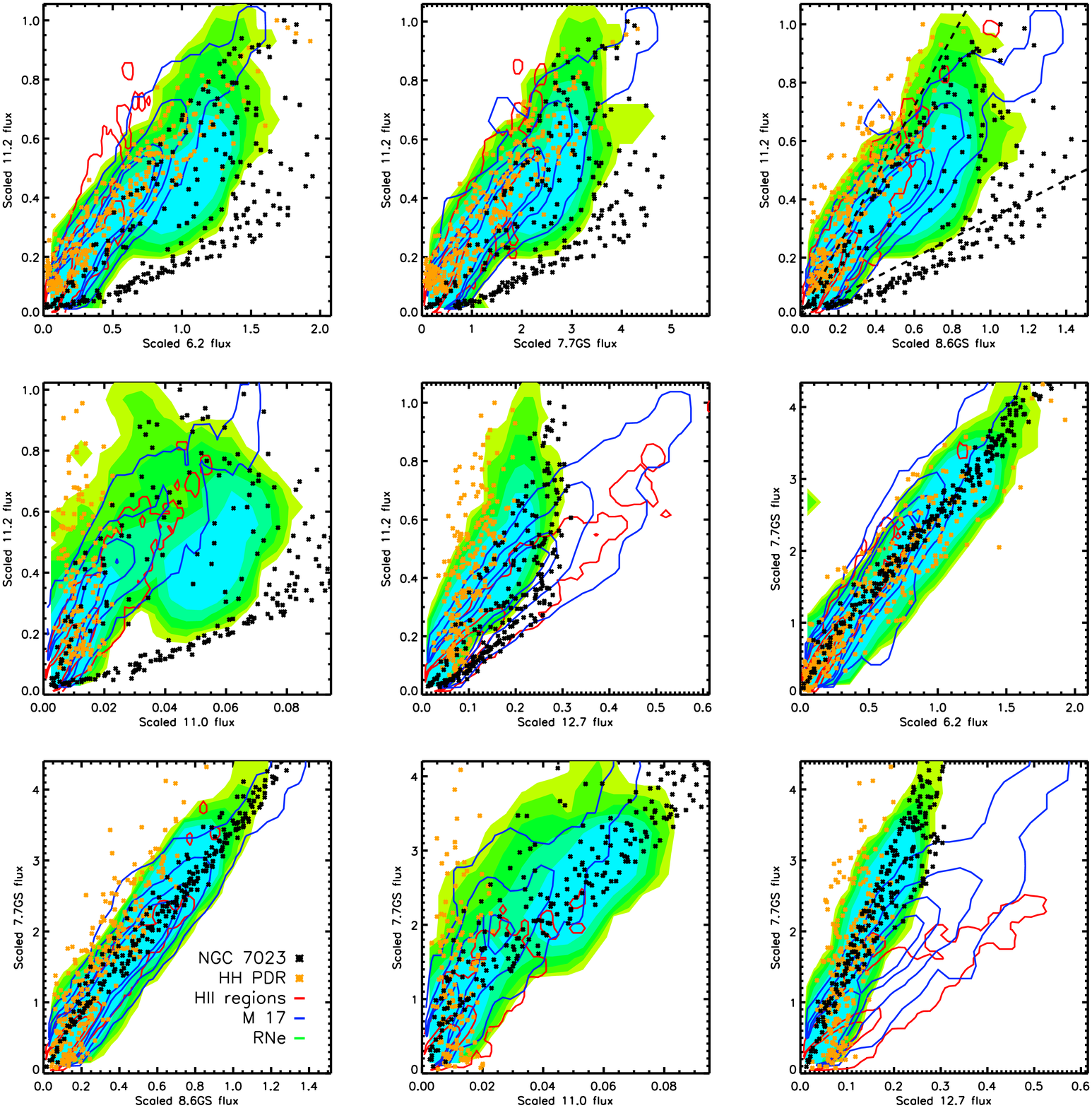}
	\end{center}
	\caption{MIR emission component flux correlations with the 11.2 \micron\ PAH band (normalized to the peak 11.2 flux of each source). For each panel we have only included points for which there is a 3-sigma detection for each axis. In the rightmost panel of the top row we include the trends found by \citet{2014ApJ...795..110B} for the NGC~7023 data (see text) as dashed lines. Symbols and contours as in Figure~\ref{fig:rat_corr}, contour levels given in Table~\ref{tables:contlevs}. }
	\label{fig:lm_corr_all}
\end{figure*}

\paragraph{Normalization} \citet{2013ApJ...769..117B} showed that there is useful information regarding local physical conditions along the line of sight encoded into the PAH band flux to PAH band flux correlation. Hence, we examine our sample in such a way as well. However, in order to visualize each of the sources in the same parameter space, we scaled the fluxes from each object, as the pixel by pixel band fluxes between the sources vary more than three orders of magnitude. The weakest and strongest emission is associated with the Horsehead PDR and W49A, respectively. Using the 11.2 \micron\ PAH band flux as an example, the W49A peak flux is 2.4 $\times$ 10$^{-18}$ W m$^{-2}$, while the brightest Horsehead 11.2 \micron\ PAH band measurement is 5.4 $\times$ 10$^{-21}$ W m$^{-2}$. In order to consider the interrelationships of the fluxes of the PAH bands, we scaled each source's fluxes by the \emph{peak} 11.2 \micron\ flux, such as to yield 11.2 \micron\ fluxes in the range of 0--1 for each source while the 7.7 \micron\ band fluxes, for example, vary from 0 -- $\sim$4. This process maintains the intrinsic relationships between each band and is very different from the usual process of normalization (discussed in the previous Section). So to summarize, the peak 11.2 \micron\ PAH flux was used as a scale factor for all of the flux measurements for each pixel of each source, rather than dividing each pixel, for example, by its own 11.2 \micron\ flux measurement as done in the traditional way. 

The peak 11.2 \micron\ band fluxes used as scaling factors are presented in Table~\ref{tables:norms}. The flux relationships between the MIR PAH emission components are presented in Figure~\ref{fig:lm_corr_all}. In these figures we have included only points for which the band was detected with a signal-to-noise ratio greater than three for both axes. 

\begin{table}
\caption{\label{tables:norms}Scaling Factors}
\begin{center}
\begin{tabular}{c c }
\hline\hline
Object            & Peak 11.2 \micron\ flux\\
                  & [$\times$ 10$^{-20}$ W m$^{-2}$] \\
\hline
\\
\textsl{H~\textsc{ii} Regions:}\\
\hline
W49A              & 241.12 \\
G37.55-0.11       & 126.14 \\
IRAS~12063-6259   & 294.22 \\
G11.94-0.62       & 225.04 \\
M~17              & 157.89 \\
Horsehead PDR     & 0.5466 \\
\\
\textit{Reflection Nebulae:}\\
\hline
NGC 2023 S        & 17.43 \\
NGC 7023          & 27.38 \\
NGC 1333          & 68.86 \\
\\
\hline\hline
\end{tabular}
\end{center}

\end{table}

\paragraph{NGC 7023} The reflection nebula NGC~7023 has been examined in terms of its flux-flux correlations by \citet{2014ApJ...795..110B}. Therefore, we will not repeat the detailed findings of that paper here, although, we will make some comparisons later on. In general though, the NGC~7023 points agree with those found for the other RNe, with the exception of an additional component in many cases. This component is defined by weak 11.2 emission and thus forms the lower trend in the top panels of Figure~\ref{fig:lm_corr_all}. In the third panel of the top row of Figure~\ref{fig:lm_corr_all}, we show the two general trends identified for NGC~7023 by \citet{2014ApJ...795..110B}, which they label the `PDR' line (upper) and the `diffuse' line (lower). Where appropriate, we will adopt this convention throughout. \citet{2014ApJ...795..110B} attributed the observed bifurcation to changes in the 11.2 \micron\ complex profile, specifically the appearance of extra emission appearing redward of 11.3 \micron\, which they refer to as the `red wing'. In general, this split is evident for all of the relationships between ionized PAH bands (at 6.2, 7.7, 8.6 and 12.7 \micron) and the neutral band at 11.2 \micron, albeit very weakly for 12.7 vs. 11.2. In addition \citet{2015ApJ...806..121B} found that the different gradients on this plot could be ascribed to different degrees of ionization, with the PDR line representing an ionized fraction of 0.4 and the diffuse correlation having an ionized fraction of around 0.7. In our Figure it can be seen that the Horsehead points seem to adopt a slightly steeper gradient than the PDR line, perhaps suggesting that the PAH molecules in the Horsehead nebula possess an ionization fraction even lower than the 0.4 found for the PDR correlation. 

\paragraph{Groupings} In general, all of the \HII\ region points follow the `PDR' component identified by \citet{2014ApJ...795..110B}, with the exception of some outliers seen for M17 and the Horsehead. There are some small inconsistencies in gradient though, with the \HII\ regions and M17 points forming slightly different gradients in the 11.2 vs. 6.2 plot, and the Horsehead points forming an almost vertical trend in the 11.2 vs. 11.0 plot. The NGC~1333, 2023 points (grouped together as `RNe') also follow the `PDR' component, albeit with a distinct bulge towards the `diffuse' component seen in the NGC~7023 points. In some cases, this bulge overlaps with the diffuse trend  (particularly when considering the 11.0 \micron\ band). In terms of the \citet{2014ApJ...795..110B} interpretation of this diagram, this would imply that a subset of the RNe points possess ionization fractions similar to that seen in NGC~7023. Finally, in the 11.2 vs. 12.7 plot, we see that the Horsehead points actually follow the RNe points much better than the \HII\ regions or M17 samples, contrary to what might be expected.

\paragraph{Trends with the 7.7 \micron\ band} The middle-right panel and bottom row of Figure~\ref{fig:lm_corr_all} feature the relationships between the 7.7 \micron\ PAH band fluxes (7.7GS) and the other PAH bands. For 6.2 and 8.6GS, all of the points are consistent with linear relationships between the bands, and for 11.0 we see what amounts to a curve, where every increase in 7.7 leads to a slightly smaller increase in 11.0. This shape is effectively the same as that seen for the 11.2 vs. 11.0 correlation (middle left panel of Figure~\ref{fig:lm_corr_all}), implying that either the 7.7 or 11.0 \micron\ band measurement features a neutral component\footnotemark. The most interesting result is that the slope of the trend in the 7.7 vs. 12.7 plot is different for each source, possibly reflecting the physical conditions. Again, this suggests that the 12.7 \micron\ PAH band is comprised of emission from both neutral and ionized PAH molecules. This idea has been suggested by several groups, particularly \citet{2014ApJ...791L..10C} who showed that both ionized and neutral PAHs with an `armchair' structure could contribute to emission at around 12.7 \micron. Subsequently \citet{2014ApJ...795..110B} found observational evidence of this effect in NGC~7023, which we show here is a general behavior.

\footnotetext{This concept is further examined using high-resolution mode \textit{Spitzer}-IRS data by Shannon, Stock \& Peeters, (2016, submitted).}

\subsection{Summary}

Our results can be summarized as follows:

\begin{itemize}
\item The \HII\ regions behave differently from the RNe. This was clearest when we consider the range of variations between the PAH ratios (Figure~\ref{fig:rat_corr}) where for the 8.6/11.2 and 11.0/11.2 ratios the RNe  span a dynamic range twice that of the \HII\ regions. We note that our \HII\ region sample is systematically more distant than the RNe sample and suspect that this compresses the dynamic range of the \HII\ region data by spatial dilution.
\item There are systematic differences between the slopes recovered for the different groups present in many of the found correlations. For example, even in the most well-known correlation 6.2/11.2 versus 7.7/11.2, the \HII\ region and M~17 samples have steeper slopes than the RNe.
\item The data suggests that the 11.0 and 12.7 \micron\ bands may arise from a combination of neutral and ionized PAH species. For the 12.7 \micron\ band this effect is very clear in Figure~\ref{fig:lm_corr_all}, where the different groups have different slopes.
\item NGC~7023 has unique PAH emission characteristics which seem to be driven by the weaker 11.2 \micron\ band emission in the diffuse region (around a factor of four weaker for a constant intensity of the 6 -- 9 \micron\ bands), which causes the traditional correlation plots to have extended ranges (by at least a factor of two) and the flux versus flux correlation plots to split into two branches (labeled diffuse and PDR). 
\end{itemize}

\section{Discussion}\label{sec:disc}

\subsection{PAH Properties in H~\textsl{\textsc{ii}} regions Regions and RNe}

In Figure~\ref{fig:lm_corr_all} we showed that the scaled flux correlations between most of the PAH bands had a variety of shapes, ranging from linear to bifurcated. The bifurcated correlations belong exclusively to NGC~7023, and this is likely due to its unique geometry with respect to our line of sight, producing a very sharp edge between two distinct PAH emitting regimes (e.g., \citealt{2013ApJ...769..117B,2014ApJ...795..110B,2015ApJ...806..121B}). The shapes seen in the scaled correlation plots for the other objects seemed to take two forms: linear, and linear plus an offset peak (this pattern is clearest in the 11.2 vs. 11.0 panel of Figure~\ref{fig:lm_corr_all}, where the offset peak is centered around $x$ = 0.06, $y$ = 0.5 for the RNe). In the following discussion we follow the naming convention of \citet{2014ApJ...795..110B} and refer to the upper linear trend as the PDR material and the lower trend as the diffuse material. In these terms then, the linear component evident for most of the objects is very close to the `PDR' line defined by \citet{2014ApJ...795..110B}, although different sources seem to have slightly different slopes. While the second pattern consists of a section of the PDR line at lower intensity levels and an offset peak for the brighter regions. We can understand this pattern as being a mixture of environments along the line of sight, with the points within the offset peak being dominated (although to what degree we cannot say) by ionized PAH spectra from regions similar to the diffuse region seen for NGC~7023, and the linear component dominated by colder material behind the PDR front (i.e. experiencing a lower radiation field similar to the PDR regions of NGC~7023). 

This observation can help us understand the seemingly uniform nature of the \HII\ region scaled flux correlations, which seem to all fall into the category of `linear' correlations, i.e. dominated by emission from the PDR surrounding the \HII\ regions. However, If we assume that the much more distant \HII\ regions might possess environments similar to the NGC~7023 `diffuse' component, then one may expect slight curvature in the linear correlations because the pixels containing the diffuse component would be diluted with normal `PDR' material to a greater extent -- with the overall effect being a slight shift rather than forming a separate peak. In this case the PDR material is still the major component of the spectrum. We can test this idea immediately by examining Figure~\ref{fig:lm_corr_all} for evidence of curvature in the `linear' trends, with the expectation that the curvature would increase (i.e. move towards the combination of PDR and diffuse components seen for the RNe) for the closer objects with pixels covering smaller spatial areas (see Table~\ref{tables:objects} for the area of 1 IRS/SL pixel in square parsecs for each object). This trend is evident if we compare the \HII\ region sample (generally very distant, dominated by W49A points at distance $>$10 kpc) and M17 (almost a factor of ten closer) in Figure~\ref{fig:lm_corr_all}, indeed the M17 trend is curved towards the RNe points (this is especially evident in the top row where 11.2 \micron\ band is plotted vs. 6.2, 7.7 and 8.6 \micron\ band measurements). Interestingly, the Horsehead points match this pattern for the 6.2GS and 7.7GS bands vs. the 11.2 \micron\ band scaled flux correlation, but not the 8.6GS vs. 11.2 \micron\ scaled flux correlation. 

\begin{figure*}
	\begin{center}
	\includegraphics[width=17cm]{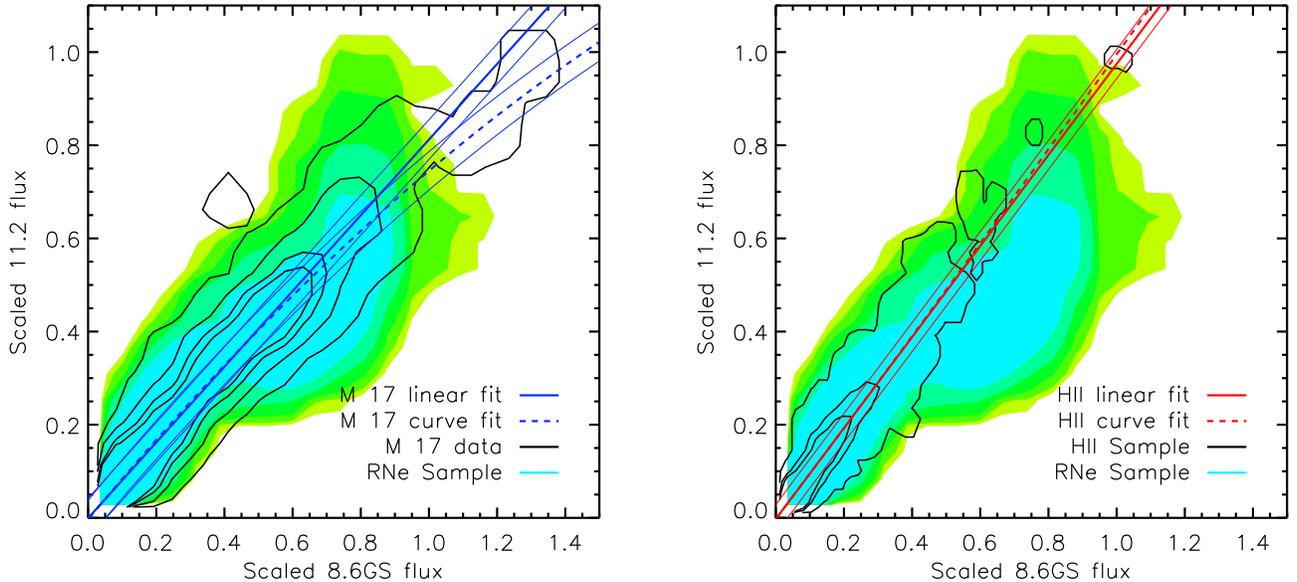}
	\end{center}
	\caption{Best fits for straight line (solid) and polynomial models of form $f(x) = a+bx+cx^2$ (dashed) for M 17 data (blue, left) and the \HII\ region sample (red, right) superimposed on the point density contours of the RN NGC~1333 and NGC~2023 from the top right panel of Figure~\ref{fig:lm_corr_all} and the point density contours for the M 17 data points and the \HII\ region data points, respectively. Thin blue and red lines correspond to the data dispersion following the prescription from \citet{2008ApJ...679..310G}. Contour levels are the same as Figure~\ref{fig:lm_corr_all} and given in Table~\ref{tables:contlevs}.}
	\label{fig:curveHII}
\end{figure*}

We show this effect in Figure~\ref{fig:curveHII}, in which we replicate the RNe contours of the 11.2 vs. 8.6 \micron\ scaled band intensity correlation plot and show best fits for the \HII\ regions and M~17 data using both straight line and polynomial ($f(x) = a+bx+cx^2$) models taking into account both $x$ and $y$ uncertainties calculated from the signal-to-noise ratios. For the \HII\ regions sample the curved fit follows the straight line almost exactly showing that there is almost no curvature, conversely for M17 a significant deviation from the straight line towards the secondary RNe peak is seen (with an accompanying increase in reduced $\chi^2$). There is a significant amount of scatter around these trends, which shows that even in pixels which have similar dispositions of environments along the line of sight, there can be significant changes in flux. Simple models like our polynomial fit, cannot capture these changes (in fact, the scatter for the Horsehead points forces the best fitting curve model to turn over entirely). However our model does seem to capture the small changes present in more distant objects like M~17. For the \HII\ regions, that are around a factor of ten more distant than M~17, the diffuse component cannot be detected in this way, as if it is present, its emission is overwhelmed by the PDR-like material dominating the beam.

Earlier we commented that the Horsehead PDR points do not seem to follow the expected pattern of curvature despite being nearby, where it would be expected that the increased spatial resolution would yield a clearer detection of any diffuse component. The likely reason for this is, again, the size of the beam used to make our map, but in the opposite sense from the other \HII\ regions -- the Horsehead region is so large on the sky that the diffuse region is not present in our map! \citet{2015A&A...576A...2O} show IRS/SL spectra for a selection of positions between the IC~434 / Horsehead PDR interface and the ionizing star (their Figure 4), in which they show that the spectrum of their region 6, several arcminutes outside of our map, seems to have a PAH spectrum reminiscent of that seen for the `diffuse' emission from NGC~7023.

\subsection{Physical Conditions and PAH Band Ratios}
 
We have seen that the observed PAH band intensity ratios can be attributed mainly to changes in the overall ionization state of the emitting PAH molecules within the beam of each observation. These feature ratios should therefore be related to the physical conditions\footnotemark\ in which the PAH molecules find themselves. Empirical calibrations between the band ratios and physical conditions have been attempted several times (e.g., \citealt{2005ApJ...621..831B}, \citealt{2008ApJ...679..310G} and \citealt{2015ApJ...806..121B}). These three studies took very different approaches though, with \citet{2005ApJ...621..831B} estimating the physical conditions in their sample based on other information (e.g., spectral types and geometry); \citet{2008ApJ...679..310G} using only well studied sources for which the physical conditions were known beforehand; and \citet{2015ApJ...806..121B} using observations to calibrate the ionization fractions found using NASA Ames PAH database \citep{2010ApJS..189..341B,2014ApJS..211....8B} with $\gamma$, using only the PAH spectra. We adopted a mixture of the first two approaches, using literature values where present, and estimating other parameters using simple relationships between, for example, the stellar spectral types, G$_0$, and continuum radio emission measurements and the nebular n$_e$. Given the very different strategy employed by \citet{2015ApJ...806..121B} to compute physical conditions, we present their results as a point of comparison as we would not necessarily expect their methodology to produce the same results as those presented here or those by \citet{2008ApJ...679..310G}.

\footnotetext{For example the radiation field strength, G$_0$ in units of the Habing field (1.2 $\times$ 10$^{-4}$ [erg cm$^{-2}$ s$^{-1}$ sr$^{-1}$]; \citealt{1968BAN....19..421H}), various permutations of the local density (e.g., n$_H$ or n$_e$ [cm$^{-3}$]) or the temperature T [K]. Sometimes these quantities are grouped together to define the ionization parameter, $\gamma$ which is equal to G$_0$/n$_e$ $\times$ (T / 1000 K)$^{0.5}$ \citep{2010pcim.book.....T}. }

Firstly, we restrict our sample to those sources where we believe that a simple estimate is physically meaningful, i.e. where we believe that the simple estimates of physical conditions correspond to the location of the dominant emitting PAH population. In some cases this is relatively straightforward, as the geometry of the system is obvious. For example, the IC~434 / Horsehead PDR front is at a known (large) distance from its exciting star and hence experiences no strong gradients in G$_0$. Conversely, objects such as M~17 are very difficult to treat in a spatially resolved way, as although there exists a prescription for the varying of G$_0$ throughout our field of view (based on the spectral types and distances of the stars in the central cluster; \citealt{2013ApJ...774L..14S}) we do not have corresponding spatially resolved measurements of density or temperature. As such, we concentrate on including the geometrically simplest sources in our sample: the UC-\HII\ regions, including those measured as part of the W49A map: the ultra-compact \HII\ regions CC and DD (see, for example, \citealt{1997ApJ...482..307D, 2014ApJ...791...99S}). 

The UC-\HII\ regions can be characterized using only radio observations of the free-free hydrogen emission continuum. Such observations typically provide us with: a) the stellar spectral type; b) the radius of the \HII\ region; and c) the hydrogen density of the \HII\ region. In order to convert these values into a measurement of the physical conditions in the PDR region, we employ the following relationships. Firstly, the density of a PDR (in terms of the number of hydrogen nuclei n$_H$, regardless of ionization or molecular state) is about a factor of 30 higher than the electron density of the ionized region (e.g., \citealt{2010pcim.book.....T}). We can convert this n$_H$ into the election density, n$_e$, of the PDR by assuming that the dominant source of electrons is the ionization of neutral carbon atoms, e.g., that n$_e$ $\simeq$ (C/H)n$_H$ $\simeq$ 1.6 $\times$ 10$^{-4}$n$_H$ if we assume galactic abundances. Here we are explicitly following the method of \citet{2008ApJ...679..310G}, although this relationship is a fundamental part of the structure of PDRs. An average value for G$_0$ can be calculated from the stellar spectral type if we know the radius of the PDR. This relationship is given by \citet{2010pcim.book.....T} as: 

\begin{equation}
G_0 = 625 \frac{L_*\chi}{4\pi d^2},
\end{equation}

where $L_*$ is the stellar luminosity, $\chi$ is the fraction of far-UV photons and $d$ is the diameter of the \HII\ region. Using the relationships provided by \citet{2003ApJ...584..797P} we can translate the stellar spectral type into the FUV luminosity $L_*\chi$, and thereby arrive at the results presented in Table~\ref{tables:physconds} for our UC-\HII\ regions. This process is performed routinely in order to calculate $G_0$ for PDRs (e.g., \citealt{1992ApJ...390..499M,2015A&A...579A..67S}). The final parameter, T, only appears in its square root, and so, variations within the small range of temperatures physically plausible for PDRs, such as those studied here, only change $\gamma$ by around 10\%, with a typical temperature being around 400 K. Therefore we adopted a value of 10\% as the uncertainty of the calculated $\gamma$ factors. For each source we calculated the 6.2/11.2 band intensity ratio by finding the average and standard deviations of the 6.2/11.2 band intensity ratios of all of the associated spectral pixels. 

\begin{table*}
\caption{\label{tables:physconds} Physical conditions for UC\HII\ regions}
\begin{center}
\begin{tabular}{c c c c c c c c}
\hline\hline
Object            & Spectral Type & PDR radius [pc] & \HII\ density [cm$^{-3}$] & log(G$_0$) & n$_{e,PDR}$ & T [K] & $\gamma$ \\
\hline
\\
W49A/CC           & O6            & 0.09            & 2700                      & 4.67        & 13         & 400   & 2270\\
W49A/DD           & O6            & 0.09            & 3100                      & 4.67        & 15         & 400   & 1977\\
G37.55-0.11       & O6            & 0.06            & 12000                     & 4.38        & 58         & 400   & 1070\\ 
IRAS~12063-6259   & O6            & 0.13            & 1000                      & 4.31        & 5          & 400   & 2746\\
G11.94-0.62       & O7.5          & 0.02            & 61000                     & 5.14        & 293        & 400   & 1192\\ 
Horsehead PDR$^a$ & O9            & 3.7             & --                        & 2           & 0.15       & 400   & 105\\
\\
\textit{\citet{2008ApJ...679..310G} Sources:}\\
\hline
NGC 2023          & B1.5          &                 &                           & 4.17        & 16         & 750   & 800 \\
NGC 7027          & WD            &                 &                           & 5.77        & 1600       & 2000  & 520 \\
Orion Bar         & O6            &                 &                           & 4.60        & 8          & 500   & 3519 \\
\\
\hline\hline
\end{tabular}
\end{center}
\bigskip
$^a$: All values from \citet{2015A&A...576A...2O}.\\

\end{table*}

\begin{figure}
	\begin{center}
	\includegraphics[width=8cm]{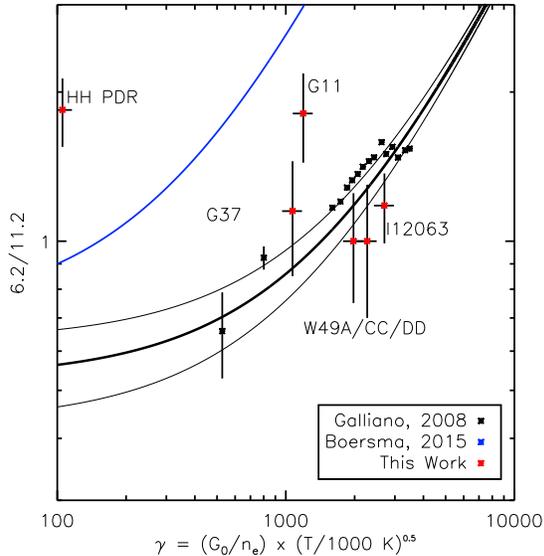}
	\end{center}
	\caption{The empirical relationship between the PAH band intensity ratios and the physical conditions (collectively referred to as $\gamma$, see main text). The black points and line refer to the work of \citet{2008ApJ...679..310G}, where the thick line is the derived linear relationship between the black points and the thinner lines represent 1$\sigma$ deviations. The red points represent the objects studied in this work, which seem to split into two groups, with four sources being consistent with the \citet{2008ApJ...679..310G} line and two sources which are not. }
	\label{fig:gal}
\end{figure}

The data from Table~\ref{tables:physconds} is shown in Figure~\ref{fig:gal} along with the points and linear regression from \citet{2008ApJ...679..310G}. Four of our UC-\HII\ regions are consistent with the \citet{2008ApJ...679..310G} data: W49A/CC, W49A/DD, G37.55-0.11 and IRAS~12063-6259. The remaining sources (G11.94-0.62, and the Horsehead) deviate from the trend, appearing above the line (by factors of 2 and 4 respectively). We first investigate possible systematics affecting our calculation of $\gamma$. The key parameter which could influence the $\gamma$ values strongly is the size of the \HII\ region as measured from the radio data. This parameter strongly affects the value of $G_0$ without changing $n_e$. For the UC-\HII\ region radii we used observations by \citet{1989ApJS...69..831W} for G11 and G37, \citet{1997ApJ...482..307D} for the W49A sources and \citet{2003A&A...407..957M} for IRAS~12063-6259. IRAS~12063-6259 is a complex source with three separate peaks in the radio map \citep{2003A&A...407..957M}. In order to include this source, we measured the approximate size of the northern peak, which shows the strongest emission and is roughly circular. For the W49A sources we did not use the radii quoted for the regions directly, as the low resolution radio observations of those regions gave very different values (a factor of three larger) than the radii given for the very similar sources (in terms of stellar spectral type and density) observed in high resolution mode. The numbers adopted and quoted in Table~\ref{tables:physconds} refer to this average of the high resolution observations of similar sources in W49A (i.e., same spectral types and densities). In order to force the G11.94-0.62 point to match the \citet{2008ApJ...679..310G} trend we need to reduce the radius of the \HII\ region by 40\%, and while \citet{1989ApJS...69..831W} do not quote uncertainties on their measurements, it is clear from inspection of their imagery (\citealt{1989ApJS...69..831W}, Figure 5a) that the emission region is larger than the 0.5\arcsec\ radius implied by a 40\% reduction in size. In addition, for the Horsehead PDR, the $\gamma$ value which was calculated by \citet{2015A&A...576A...2O}, is even more strongly discrepant, with no similar arguments to be made regarding size, temperature or $G_0$ -- so it is clear that some objects can strongly deviate from the \citet{2008ApJ...679..310G} trend.

This idea is further demonstrated by the \citet{2015ApJ...806..121B} calibration of the PAH ratio and $\gamma$ (shown in Figure~\ref{fig:gal} as a blue line). The relationship found by \citet{2015ApJ...806..121B} clearly does not match any of the points used by \citet{2008ApJ...679..310G} or from our \HII\ regions sample. That said, the Horsehead and G11 are closer to the \citet{2015ApJ...806..121B} relationship than the \citet{2008ApJ...679..310G} fit. \citet{2015ApJ...806..121B} attribute this to differences in the decomposition of the bands, i.e. that \citet{2008ApJ...679..310G} include the 11.0 \micron\ band flux in their 11.2 \micron\ band measurement. However, our decomposition does differentiate between the two bands, and we find that the 11.0 \micron\ band is on average around 5--10\% of the strength of the 11.2 \micron\ band (e.g.,~Figure~\ref{fig:rat_corr}), which is clearly not enough to explain the difference in Figure~\ref{fig:gal}, which is about 50\%. The elevated NGC~7023 PAH band ratios found by \citet{2015ApJ...806..121B} are consistent with those found for NGC~7023 in this paper (and largely at greater 6.2/11.2 values than we have shown in Figure~\ref{fig:gal}). This observation leaves us with two options: either the physical conditions in NGC~7023 are far more extreme than predicted by \citet{2015ApJ...806..121B}; or that the relationships between the PAH ratios and physical conditions are more complex than that expressed by $\gamma$. Given that other sources in our sample, noticeably the Horsehead PDR with its extremely low radiation field strength, do not fall on the \citet{2008ApJ...679..310G} trend, it seems that the second scenario is the most likely.

\section{Summary and Conclusions}\label{sec:cs}

We have reduced and analyzed a large sample of \textit{Spitzer}-IRS SL spectral cubes of \HII\ regions and RNe, primarily in terms of their PAH emission. Each spectral pixel of each cube was processed using the same spline method with the same parameters, creating a uniform sample of measurements of the PAH bands and the other emission features between 5-15 \micron. Several of these cubes have been investigated before (e.g., W49A, NGC~2023, NGC~7023) and these datasets were used as a framework to understand the results for the other objects. We then attempted to understand the flux/flux correlation plots in terms of the average environment along the line of sight and showed that the observed shapes in the correlation plots represent different mixtures of diffuse and PDR environments. Finally we investigated the relationship between the observed PAH band ratios and the physical environments of our sources.

Our main findings are:

\begin{enumerate}
\item The PAH emission from \HII\ regions is distinct from that of RNe in terms of the relationships between the band strengths.
\item We find that following general groups seem to have very similar PAH properties -- the \HII\ regions (W49A, G11.94-0.62, G37.55-0.11 and IRAS~12063-6259), the RNe (NGC~1333, NGC~2023) -- and that some sources do not fall into either of these categories (NGC~7023, the Horsehead PDR and M17).
\item Traditional correlation plots reveal that the RNe have a greater dynamic range for the PAH band ratios than H II-regions. In particular NGC~7023 which, due to the noted sharp divide between environments, has a strong reduction in 11.2 \micron\ band emission across approximately half of the map. For the other RNe the variations are still only around half of those seen for NGC~7023. This effect may be due to the fact that the \HII\ regions are systematically more distant than the RNe and, as such, the contents of each \HII\ region spectrum consist of the average of a wider variety of environments, especially as the physical area encompassed in each pixel increases.
\item Both the traditional, and the flux versus flux correlation plots reveal that the 11.0 and 12.7 \micron\ bands behave differently from the other PAH bands, in a manner consistent with being generated by a mixture of both neutral and ionized PAH species.
\item The split seen between a `diffuse' and a `PDR' component observed by \citet{2013ApJ...769..117B} in flux versus flux correlation plots is only seen strongly for NGC~7023, although the nearby \HII\ regions give hints that there may be some diffuse component contribution to their spectra.
\item The presence of a subtle diffuse component can be deduced in some sources by examining the curvature in the distribution of data points for the flux versus flux correlations between a primarily ionized band and a primarily neutral band, e.g., the 11.2 \micron\ band versus the 8.6 \micron\ band.
\item Our UC-\HII\ region sample is consistent with the \citet{2008ApJ...679..310G} relationship between physical conditions and PAH ratios, however they are inconsistent with the \citet{2015ApJ...806..121B} relationship found in a spatially resolved study of NGC~7023. This seems to indicate that there may not be one simple relationship between all of the physical conditions (density, temperature and radiation field intensity) and the PAH band ratios.
\end{enumerate}

\section*{Acknowledgments}
DJS thanks Christiaan Boersma, Jesse Bregman and the anonymous referee for their constructive comments on the manuscript. 

DJS and EP acknowledge support from an NSERC Discovery Grant and an NSERC Discovery Accelerator Grant. WDYC and SS acknowledge support from NSERC Undergraduate Student Research Awards.

LJA is grateful for an appointment at NASA Ames Research Center through  the Bay Area Environmental Research Institute (NNX14AG80A).

Studies of interstellar chemistry at Leiden Observatory are supported through advanced-
ERC grant 246976 from the European Research Council, through a grant by the Dutch
Science Agency, NWO, as part of the Dutch Astrochemistry Network, and through the
Spinoza premie from the Dutch Science Agency, NWO.

This work is based on observations made with the \textit{Spitzer Space Telescope}, which is operated by the Jet Propulsion Laboratory, California Institute of Technology under a contract with NASA.

This research has made use of NASA's Astrophysics Data System Bibliographic Services; the SIMBAD database, operated at CDS, Strasbourg, France; APLpy, an open-source plotting package for Python hosted at http://aplpy.github.com; and the IDL Astronomy Library \citep{1993ASPC...52..246L}.

\bibliographystyle{apj}
%\bibliography{paper}

\end{document}